%% file: main.tex
\documentclass[twocolumn,showpacs,%
nofootinbib,aps,superscriptaddress,%
eqsecnum,prd,notitlepage,showkeys,10pt]{revtex4-1}
\usepackage[utf8]{inputenc}
\usepackage{amssymb}
\usepackage{amsmath}
\usepackage{amsfonts}
\usepackage{amsthm}
\usepackage{graphicx}
\usepackage{dcolumn}
\usepackage{hyperref}
\usepackage{stmaryrd}
\usepackage{makeidx}
\usepackage{array}
\usepackage{subfiles}
\usepackage{chngcntr}

\makeatletter
\newcommand{\fixed@sra}{$\vrule height 2\fontdimen22\textfont2 width 0pt\shortrightarrow$}
\newcommand{\shortarrow}[1]{%
	\mathrel{\text{\rotatebox[origin=c]{\numexpr#1*45}{\fixed@sra}}}
}
\makeatother

\providecommand{\keywords}[1]
{
	\small	
	\textbf{\textit{Keywords---}} #1
}

\newtheorem{axiom}{Axiom}
\newtheorem{thrm}{Theorem}        % theorem number
\newtheorem{lem}{Lemma}        % lemma number
\newtheorem{corl}{Corollary}
\newtheorem{prop}{Proposition} % proposition number
\newtheorem{defn}{Definition}  % definition numbers 
\newtheorem{exmp}{Example}

\counterwithout{equation}{section}

\begin{document}
	\title{The Signatures of Ideal Flow Networks}
	\author{Kardi Teknomo}
	\affiliation{Petra Christian University, Indonesia \\
		University of the Philippines, Philippines}
	\date{\today}
	
	\subfile{abstract.tex}

	\maketitle
	\subfile{introduction.tex}
	\subfile{relatedWork.tex}
	\subfile{IFN.tex}
	\subfile{canonicalCycles.tex}
	\subfile{assignmentOperator.tex}

	\subfile{example_assign_operator.tex}
	\subfile{mergingOperator.tex}
	\subfile{example_merging_operator.tex}
	\subfile{signature.tex}
	\subfile{premier.tex}
	\subfile{composition.tex}
	\subfile{decomposition.tex}

\subfile{example1.tex}
	\subfile{StringAnalysis.tex}
	\subfile{irreducibilityCondition.tex}
	\subfile{conclusion.tex}

\subfile{supplementary.tex}

	\bibliographystyle{plain}
	\bibliography{main.bbl}
%	\bibliographystyle{IEEEtran}
%	\bibliography{IEEEabrv,signatureIFN.bbl}

\end{document}

%% file: abstract.tex
\begin{abstract}
	An Ideal Flow Network (IFN) is a strongly connected network where relative flows are preserved (irreducible premagic matrix). IFN can be decomposed into canonical cycles to form a string code called network signature. A network signature can be composed back into an IFN by assignment and merging operations. Using string manipulations on network signatures, we can derive total flow, link values, sum of rows and columns, and probability matrices and test for irreducibility.
\end{abstract}

\keywords{canonical cycles, network signature, composition,  decomposition, premagic}

%% file: introduction.tex
\section{Introduction}

Network theory is a fundamental area of study within discrete mathematics and has broad applications across various scientific and engineering disciplines. Traditional approaches to network analysis focus on connectivity, path finding, and flow optimization. A new and novel framework of network signatures is introduced in this paper based on the concepts of canonical cycles, offering a new representation on integer ideal flow matrix and its directed graph as summation of its canonical cycles. The network signature is a string codes for representing matrices and networks. This representation offers a bidirectional correspondence between operations on network signatures and changes in their matrix or network counterparts. Our objective is to offer a comprehensive framework by establishing the relationship of the signatures to ideal flow matrices and their properties.

This work is organized as follows: We begin by showing related works, defining the foundational elements of IFNs, including the concepts of canonical cycles and network signatures. We then present the key theorems related to the composition and decomposition of IFNs. Finally, we discuss the string manipulations to derive matrix values, total flow, row sums, column sums, link flow values, and stochastic matrices.

%% file: relatedWork.tex
\section{Related Work}

Ren\'e Descartes in 1637 \cite{descartes1637discourse} introduced Cartesian coordinates in his work "La G\'eom\'etrie," which was part of his larger work "Discourse on the Method". This system allowed geometric problems to be solved algebraically using coordinates, revolutionizing mathematics by laying the foundation for analytic geometry. The Cartesian coordinate system has since become fundamental in various fields of science and engineering, providing a new way to describe geometric objects and solve geometric problems. Similar to introduction of Cartesian coordinates as new representation of geometrical object, the network signature has the potential to transform linear algebra and graph theory through its new string-based representation.

Network flow theory has been extensively studied, with foundational work by Ahuja et al. \cite{ahuja1993network} providing a comprehensive overview of network flow algorithms and applications. The detection of elementary circuits in directed graphs, as discussed by Johnson \cite{johnson1975finding}, and the identification of strongly connected components using Tarjan's algorithm \cite{tarjan1972depth}, form the basis for understanding cycles in network structures. Canonical cycles within the Ideal Flow Network can be determined using algorithms such as those proposed by Johnson \cite{johnson1975finding} and Tarjan \cite{tarjan1972depth}. Johnson's algorithm efficiently finds all elementary circuits in a directed graph using depth-first search and backtracking with time complexity \( O((V + E) \cdot (C + 1)) \), where \( V \) is the number of vertices, \( E \) is the number of edges, and \( C \) is the number of cycles and space complexity \( O(V + E) \). Tarjan's Algorithm, on the other hand, identifies all strongly connected components in a directed graph, which can subsequently be used to detect cycles. The time complexity is \( O(V + E + C) \) and the space complexity is \( O(V + E) \). Bang-Jensen and Gutin \cite{bang2008digraphs} provide a thorough examination of digraphs, including cycle detection and canonical forms. The concepts presented in this paper build upon these foundational works, introducing new methodologies for network composition, decomposition and string representation.

A few papers provide valuable insights into network representation, canonical forms, and signature schemes. Gornitskii \cite{gornitskii2015essential} and Smith \cite{ieee1098584} discuss simplifying complex structures and canonical forms, similar to network signatures but focused on algebraic structures and control systems. Deng \emph{et al} \cite{deng2020network} and Zhang \emph{et al} \cite{zhang2023network} offer methods for embedding networks, emphasizing machine learning and attention mechanisms. Boneh \emph{et al} \cite{boneh2008signing} presents secure data representation techniques. However, none specifically address canonical cycles, network signatures, and Ideal Flow Networks, highlighting the originality of the proposed research.

%% file: IFN.tex
\section{Ideal Flow Network}

\begin{defn}
	(Strongly Connected Network): A directed graph \( G = (V, E) \) is said to be \emph{strongly connected} if for every pair of vertices \( u, v \in V \), there exists a directed path from \( u \) to \( v \) and from \( v \) to \( u \). 	
	
	\begin{equation}
	\begin{split}
	G \text{ is strongly connected} \iff \\
	\forall u, v \in V(G), \exists \text{ a path from } u \text{ to } v
	\end{split}
	\end{equation}
	
\end{defn}

\begin{defn}
	(Irreducible Matrix): A square matrix \( A \) is called \emph{irreducible} if and only if, for every pair of indices \( i \) and \( j \), there exists a positive integer \( k \) such that the \( (i,j) \)-th entry of the matrix \( A^k \) is positive.
	\begin{equation}
	\begin{split}
		A \text{ is irreducible} \iff \\
		\forall i, j, \exists k \in \mathbb{N} \text{ such that } (A^k)_{ij} > 0
	\end{split}
	\end{equation}
\end{defn}

The adjacency matrix \( \mathbf{A} \) of a directed graph \( G \) is irreducible if and only if the graph \( G \) is strongly connected. The proof can be found in \cite{seneta1973nonnegative}. This means that 
\begin{equation}
	\mathbf{A} \text{ is irreducible} \iff G(\mathbf{A}) \text{ is strongly connected},
\end{equation}
where \( G(\mathbf{A}) \) denotes the directed graph represented by the adjacency matrix \( \mathbf{A} \).

\begin{prop}\label{prop-Matrix-Irreducibility-Test}
	(Irreducibility Test): A square matrix \(\mathbf{A}\) is irreducible if and only if \((\mathbf{I} + \mathbf{A})^{n-1} > 0\), where \(n\) is the number of rows in the matrix.
\end{prop}

The proof can be found in \cite{horn2013matrix}.

\begin{defn}
	(Premagic Matrix \cite{teknomo19premagic}): A \emph{premagic} matrix is one where the sum of elements in each of its rows equals the sum of elements in the corresponding columns. 	
\end{defn}

\begin{defn}\label{defn-IFN}
	(Ideal Flow Network \cite{teknomo18MC}): An \emph{Ideal Flow Network} (IFN) is a directed graph \( G = (V, E) \) where \( V \) is the set of vertices and \( E \) is the set of edges, such that the network is strongly connected (i.e., its adjacency matrix is irreducible) and the relative flow along each edge is preserved. The sum of the weights in each row equals the sum of the weights in the corresponding column (premagic).
\end{defn}

\begin{defn}\label{defn-equivalentIFN}
	(Equivalent Ideal Flow Networks): Let \(N_1\) and \(N_2\) be two ideal flow networks. The two IFNs are called equivalent if and only if each corresponding link flow of one network is a multiple of the other network by a positive global scaling factor \(\zeta\).
	\begin{equation}
	N_1 \equiv N_2 \iff N_1 = \zeta N_2, \quad \zeta > 0 
	\end{equation}
\end{defn}

\begin{prop}
	(Scaling IFN): Multiplying the link probability by the total flow \(\kappa\) produces a new equivalent ideal flow matrix with the same total flow \(\kappa\).
\end{prop}

\begin{proof}
By definition, an ideal flow network (IFN) represents relative flows between nodes. The total flow \(\kappa\) acts as a global scaling factor for the network. Let \(P_{pq}\) be the link probability derived from 
\begin{equation}\label{eq-Link-Probability}
	P_{pq} = \frac{f_{pq}}{\kappa}.
\end{equation}  
The link flow value \(f_{pq}\) can be obtained by multiplying the link probability \(P_{pq}\) by the total flow \(\kappa\).
\( f_{pq} = P_{pq} \cdot \kappa \)
Multiplying the link probability \(P_{pq}\) by \(\kappa\) yields the original link flow value \(f_{pq}\). Hence, the new equivalent ideal flow matrix retains the same total flow \(\kappa\).
\end{proof}

\begin{corl}
    (Scaling of IFN): Multiplying the link probability by the total flow \(\kappa\) produces a new equivalent ideal flow matrix with the same total flow \(\kappa\).
    \begin{equation}\label{eq-equivalent-IFN}
        f'_{pq} = P_{pq} \cdot \kappa = f_{pq} 
    \end{equation} 
\end{corl}
\begin{proof}
The link probability \(P_{pq}\) is derived from Equation \ref{eq-Link-Probability}.
Multiply \(P_{pq}\) by \(\kappa\) to obtain the original link flow value \(f_{pq}\).
Hence, the new equivalent ideal flow matrix retains the same total flow \(\kappa\).
\end{proof}

\begin{corl}
	Equivalent IFNs have the same link probability matrix.
\end{corl}
\begin{proof}
	Let \(N_1\) and \(N_2\) be two equivalent IFNs such that \(N_1 \equiv N_2\).By definition, \(N_1 = \zeta N_2\) for some \(\zeta > 0\). The link flows of \(N_1\) are scaled versions of the link flows of \(N_2\): \( f'_{pq} = \zeta f_{pq} \).
	The link probability \(P_{pq}\) is defined as the ratio of the link flow to the total flow:  \( P_{pq} = \frac{f_{pq}}{\kappa} \).	For equivalent IFNs, the total flow \(\kappa\) is scaled by \(\zeta\), and the link flows are also scaled by \(\zeta\):   \( P'_{pq} = \frac{f'_{pq}}{\zeta \kappa} = \frac{\zeta f_{pq}}{\zeta \kappa} = \frac{f_{pq}}{\kappa} = P_{pq} \). Therefore, equivalent IFNs \(N_1\) and \(N_2\) have the same link probability matrix.
\end{proof}

\begin{prop}
	(Integer IFN): Every Ideal Flow Network can be transformed into an Integer Ideal Flow Network (Integer IFN).
\end{prop}

\begin{proof}
	Let \(\mathbf{F}\) be an \(n \times n\) matrix representing an Ideal Flow Network (IFN). The entry \(F_{ij}\) denotes the flow from node \(i\) to node \(j\). The matrix \(\mathbf{F}\) satisfies the premagic condition if
	
	\begin{equation}
	\sum_{j=1}^{n} F_{ij} = \sum_{j=1}^{n} F_{ji} \quad \forall i \in \{1, 2, \ldots, n\}.
	\end{equation}
	
	This condition ensures that the sum of flows entering and leaving any node is balanced.
	
	The node weights in the IFN, represented by the vector \(\mathbf{w} = (w_1, w_2, \ldots, w_n)\), are derived from the stationary distribution of the associated Markov Chain, as given in \cite{teknomo18MC}. These weights \(w_i\) are rational numbers, expressed as
	
	\begin{equation}
		w_i = \frac{p_i}{q_i},
	\end{equation}

	where \(p_i\) and \(q_i\) are integers, and \(q_i \neq 0\). 
	
	To convert the rational weights into integers, we compute the least common multiple (LCM) of the denominators \(q_i\):
	
	\begin{equation}
	\text{LCM}(q_1, q_2, \ldots, q_n) = L.
	\end{equation}
	
	By scaling each weight \(w_i\) by \(L\), we obtain integer weights:
	
	\begin{equation}
	w_i' = L \cdot w_i = \frac{L \cdot p_i}{q_i} = \frac{L}{q_i} \cdot p_i,
	\end{equation}
	
	resulting in the integer ideal flow matrix \(\mathbf{F'}\).
\end{proof}

The link weights in an Ideal Flow Network (IFN) represent relative flows that are preserved throughout the network, ensuring that the sum of flows entering and leaving any node is balanced. This property is referred to as premagic. The node weights in an IFN are derived from the stationary distribution of its associated Markov Chain \cite{teknomo18MC}. Since these weights are expressed as relative flows, they are rational numbers. By determining the least common multiple (LCM) of the denominators of these rational numbers, we can scale all weights to transform them into an integer ideal flow matrix.

%% file: canonicalCycles.tex
\section{Canonical Cycle Operations}
	
	While the nodes in the cycle's node sequence may bear arbitrary labels, these labels can be standardized into canonical forms. 
	
	\begin{defn}
		(Canonical Cycle): A \emph{canonical cycle} \(\hat{c}\) is a sequence of nodes, denoted by lowercase letters, which starts at a node and follows a path through other nodes without repetition, eventually returning to the starting node. The canonical representation omits the end node, as it is identical to the start node.
	\end{defn}
	
	A canonical cycle is a cycle in a directed graph where the smallest node (according to some ordering) is the starting point of the cycle.
	
	\begin{prop}
		(Existence of Canonical Cycles in IFN): There exists at least one canonical cycle in an Ideal Flow Network.
	\end{prop}
	
	\begin{proof}
	Every strongly connected directed graph contains at least one cycle. Therefore, an IFN, being strongly connected, will contain at least one canonical cycle.
	\end{proof}
	
	Since an IFN must be a strongly connected network, it necessarily contains at least one cycle due to the properties of strong connectivity. Specifically, in any strongly connected directed graph, there exists at least one cycle.

%% file: assignmentOperator.tex
\subsection{Assignment Operator}

The assignment of a term generates an adjacency list where each consecutive pair of nodes in the cycle is assigned a value equivalent to the term's coefficient.

\begin{defn}
	(Term) A \emph{term} is a pair \((\alpha, \hat{c})\) where \( \alpha \) is the coefficient and \( \hat{c} \) is the canonical cycle.
\end{defn}

\begin{defn}
	(Coefficient): A \emph{coefficient} of a term \( \alpha \) is an integer that represents the number of times a canonical cycle \( \hat{c} \) is assigned to the network.
\end{defn}

\begin{defn}
	(Assign Operator): The \emph{assign} operator on a cycle \( \hat{c} \) with coefficient \(\alpha\) is defined as the operation that adds \(\alpha\) units of flow along the canonical cycle \( \hat{c} \) in the network. If a node or link in the cycle does not exist in the network, it is added. 
\end{defn}

Given a canonical cycle \( \hat{c} = (v_1, v_2, \ldots, v_k, v_1) \), the assignment operator is:

\begin{equation}
B = A \cdot (\alpha, \hat{c})
\end{equation}

where \( B \) is the adjacency list after assigning \(\alpha\) units of flow. For each node \( v_i \) in \( \hat{c} \), the updated adjacency list is:

\begin{equation}
B(v_k, v_{k+1}) = A(v_, v_{k+1}) + \alpha 
\end{equation}

if the link \( (v_k, v_{k+1}) \) is in \( A \), otherwise it is initialized to \( \alpha \).

Initially, adjacency list  \( A \) is an empty list. For each node \( v_k \) in \( \hat{c} \):

\begin{itemize}
	\item If \( v_k \) is not in \( A \), add \( v_k \) to \( A \).
	\item If the link \( (v_k, v_{k+1}) \) is not in \( A \), add the link \( (v_k, v_{k+1}) \) to \( A \) with weight 0.
	\item Increment the weight of the link \( (v_k, v_{k+1}) \) by \(\alpha\).
\end{itemize}

%% file: example_assign_operator.tex
 % example assign operator
 \begin{exmp}
 Suppose we have a term \((\alpha, \hat{c})\) = \((3, abcd\) then the adjacency list is built and expanded using the assign operator is
\[
\begin{split}
A = \{ (a \rightarrow b: 3), (b \rightarrow c: 3), (c \rightarrow d: 3), (d \rightarrow a: 3) \} 
\end{split}
\]
\end{exmp}

%% file: mergingOperator.tex
\subsection{Merging Operator}

The merging operation combines multiple adjacency lists into a single adjacency list.

\begin{defn} 
	(Merge Operator): The \emph{merge} operator denoted by \(+\) combines two networks by adding the weights of the links. 
\end{defn}

Given two adjacency lists \( A_1 \) and \( A_2 \), the merged network is:
\begin{equation}
A_1 + A_2 = A_3
\end{equation}
where \( A_3 \) is the resulting adjacency list.
\begin{itemize}
\item For each node \( v \) in \( A_1 \cup A_2 \):
\item For each link \( (v, u) \) in \( A_1 \cup A_2 \):
\item The weight of the link \( (v, u) \) in \( A_3 \) is the sum of the weights of the link in \( A_1 \) and \( A_2 \).
\end{itemize}

The updated adjacency list \( A_3 \) is given by:
\begin{equation}
\begin{split}
A_3(v, u) = (A_1(v, u) \text{ if } (v, u) \in A_1 \text{ else } 0) + \\
(A_2(v, u) \text{ if } (v, u) \in A_2 \text{ else } 0)
\end{split}
\end{equation}

%% file: example_merging_operator.tex
\begin{exmp}
      The adjacency list \(A_1\) is for term \((\alpha_1, \hat{c}_1)\) = \((3, abcd)\) and adjacency list \(A_2 = (\alpha_2, \hat{c}_2)=  (2, cdae) \).
      \[
      \begin{split}
      A_1 = \{ (a \rightarrow b: 3), (b \rightarrow c: 3), (c \rightarrow d: 3), (d \rightarrow a: 3) \}\\
      A_2 = \{ (a \rightarrow e: 2), (c \rightarrow d: 2), (d \rightarrow a: 2), (e \rightarrow c: 2) \}
      \end{split}
      \]
      The adjacency list is built and expanded using the merging operator for \( A_1 + A_2 \):
      \[
      \begin{split}
      A_3 = \{ (a \rightarrow b: 3, a \rightarrow e: 2), (b \rightarrow c: 3), (c \rightarrow d: 5),  \\
            (d \rightarrow a: 5),(e \rightarrow c: 2) \} 
      \end{split}
      \]
\end{exmp}

%% file: signature.tex
\section{Signature from Canonical Cycle}

We can represent the assignment and merging operations discussed in previous section in new way which is simpler and more novel. The network signature is a string code formed by the sorted canonical cycles that build the network based on assign and merging operations. Each cycle in the network signature is associated with an integer coefficient representing the number of times the cycle is assigned to the network.

\begin{axiom}
	The coefficient \(\alpha_i\) of a term in a cycle network signature linearly affects the flow value and related computations.
\end{axiom}

\begin{defn}
	(Network Signature): a \emph{network signature} is a string representation involving the summation of terms, where each term is expressed as the product of a coefficient and a canonical cycle. Formally, a network signature is a sum of terms:
	\begin{equation}
	\mathbf{F} = \sum_{i=1}^{n} \alpha_{i} \hat{c}_{i} 
	\end{equation} 
	where \( \alpha_i \) are the coefficients and \( \hat{c}_i \) are the canonical cycles.
\end{defn}

\begin{prop}	
	(Non-uniqueness of Signatures): Network signatures for the same ideal flow matrix are not unique; there exist multiple equivalent signatures.
\end{prop}
\begin{proof}
	Consider two signatures \(S_1\) and \(S_2\) that represent the same ideal flow matrix. \(S_1\) and \(S_2\) can be permuted or rearranged without altering the matrix representation. Hence, different sequences or combinations of cycles and coefficients can yield the same flow matrix.
\end{proof}	

\subsection{Identical and Equivalent Signature}

\begin{defn}
(Identical signatures):	Two network signatures are called \emph{identical signatures} if and only if they yield the same adjacency matrix and utilize identical canonical cycles.
\end{defn}

\begin{exmp}
The signatures \(bca + 2cab\) and \(3abc\) are identical because they use the same canonical cycle \(abc\) in different permutations and produce the same matrix.
\end{exmp}

\begin{defn}
(Equivalent signatures): Two network signatures are called \emph{equivalent signatures} if and only if they yield the same adjacency matrix but utilize different canonical cycles.
\end{defn}

\begin{exmp}
	Consider the following two equivalent network signatures: \(a + abcd + 3b + bd\) and \(3b + a + bcd + abd\). Both of these signatures yield the same adjacency matrix:
	\[
	\mathbf{F} = 
	\begin{bmatrix}
	1 & 1 & 0 & 0 \\
	0 & 3 & 1 & 1 \\
	0 & 0 & 0 & 1 \\
	1 & 1 & 0 & 0 
	\end{bmatrix}
	\]
	The terms are different, but the flow through the network is exactly the same.
\end{exmp}

Thus, it is possible for different network signatures, whether they use identical or equivalent cycles, to produce the same matrix. This demonstrates the non-uniqueness of network signatures.	

%% file: premier.tex
\subsection{Premier Network}
	
The ideal flow matrix is called a premier network if the weight vector \( \mathbf{x} = [\alpha_i] \) is equal to the vector of ones, indicating that all possible canonical cycle terms has a coefficient of one.

\begin{defn}
	(Premier Network): A \emph{premier network} is an IFN, where all possible canonical cycles are assigned exactly once. The premier network is denoted by \( F^* \) has network signature
	\begin{equation}
		\mathbf{F^*} = \sum_{i=1}^{n}  \hat{c}_{i} 
	\end{equation} 
\end{defn}

A premier network is a special case of an IFN where the coefficients of the network signature are all ones and it includes all possible canonical cycles. This means each canonical cycle is assigned exactly once.

%% file: composition.tex
\subsection{Composition}

\begin{defn}
	(composition): A \emph{composition} refers to the transformation of a network signature into an adjacency matrix, achieved through the operations of assignment and merging.	
\end{defn}

To compose an Ideal Flow Network (IFN) from a network signature, the algorithm proceeds as follows:
\begin{enumerate}
\item \emph{Initialization}: Start with an empty adjacency list.
\item \emph{Assignment and Merging}: For each term in the signature, perform the assignment operation by adding the coefficient value to the corresponding node sequence. Merge the resulting adjacency lists iteratively.
\item \emph{Conversion}: Convert the final adjacency list into the corresponding adjacency matrix \(\mathbf{F}\).
\end{enumerate}

Alternatively, we can also use linear equation as follow. IFN composition is the process of reconstructing the link flow vector \( \mathbf{y} \) given the link-cycle matrix \( \mathbf{H} \) and the weight vector \( \mathbf{x} \). This can be formulated as a linear equation:
\begin{equation}
\mathbf{y} = \mathbf{H} \mathbf{x}
\end{equation}

%% file: decomposition.tex
\subsection{Decomposition}

\begin{defn}
	(Decomposition): A \emph{decomposition} denotes the transformation of an ideal flow matrix into a network signature. 
\end{defn}

Algorithmically, this decomposition process can be conducted by assigning a negative of the minimum flow as the coefficient for each term during the assignment operation. To decompose an Ideal Flow Network (IFN) into a network signature, the algorithm proceeds as follows:
\begin{enumerate}
\item \emph{Cycle Identification}: Begin at any starting node and trace the node sequence until no more nodes can be traversed, thereby identifying a canonical cycle.
\item \emph{Minimum Flow Determination}: Determine the minimum flow along the identified node sequence. This minimum flow becomes the coefficient of the corresponding term in the signature.
\item \emph{Flow Deduction}: Subtract the minimum flow from the identified node sequence in the network by assignment operator with coefficient of negative minimum flow.
\item \emph{Iteration}: Repeat the canonical cycle identification, minimum flow determination, and flow deduction steps until the network is devoid of flow.
\item \emph{Termination}: The process terminates when the network has no remaining flow, resulting in the complete network signature.
\end{enumerate}

Alternatively, we can also use linear equation as follow. Given an ideal flow matrix \( \mathbf{F} \), we utilize Tarjan's algorithm to identify all cycles within the network. Using the detected cycles, we construct the link-cycle matrix \( \mathbf{H} \), where the rows correspond to the links and the columns correspond to the cycles.

The process of IFN decomposition involves determining the coefficients of each cycle term. This is equivalent to finding the weight vector \( \mathbf{x} = [\alpha_i] \) from the ideal flow matrix \( \mathbf{F} \). The steps are as follows:

\begin{enumerate}
\item Form the Link Flow Vector: Construct the link flow vector \( \mathbf{y} \), representing the flow along each link in the network.
\item Form the Link-Cycle Matrix: Construct the link-cycle matrix \( \mathbf{H} \) based on the cycles identified using Tarjan's algorithm.
\item Solve the Linear System: Use the generalized inverse or least squares method to solve the integer linear system:
\begin{equation} 
\mathbf{x} = \mathbf{H}^+ \mathbf{y}
\end{equation}
where \( \mathbf{H}^+ \) is the generalized inverse of \( \mathbf{H} \).
\end{enumerate}

%% file: example1.tex
\begin{exmp}
We will show the comprehensive example for the above concept. Let us start with the following adjacency matrix. 

\[
\mathbf{A} = \begin{array}{c|cccc}
& a & b & c & d \\
\hline
a & - & 1 & - & - \\
b & - & - & 1 & - \\
c & 1 & - & - & - \\
d & - & 1 & - & - \\
\end{array}
\]

All possible canonical cycles of the adjacecy matrix are \(\hat{c}_1=abcd\) and \(\hat{c}_2=abc\).

To make the adjacency matrix into an ideal flow matrix, we set the premagic properties by setting it into variable form:

\[
\mathbf{F} = \begin{array}{c|cccc|c}
& a & b & c & d & \Sigma \\
\hline
a & - & x+y & - & - & x+y \\
b & - & - & x+y & - & x+y \\
c & y & - & - & x & x+y \\
d & - & x & - & - & x \\
\hline
\Sigma & x+y & x+y & x & x & \kappa \\
\end{array}
\]

Connecting the entries of the adjacency matrix and the equation forms of the flow matrix, we have the following flow equations for each link:

\[
\begin{aligned}
ab: & \quad x+y \geq 1 \\
bc: & \quad x+y \geq 1 \\
ca: & \quad y \geq 1 \\
cd: & \quad x \geq 1 \\
da: & \quad x \geq 1 \\
\end{aligned}
\]

The simplest solution to satisfy all the constraints above is to set \(x=1\), \(y=1\). 

This is the same as assigning cycle \(\hat{c}_1=abcd\) and \(\hat{c}_2=abc\) once, making it a premier IFN.

\[
\mathbf{F}^* = abcd + abc
\]

\[
\mathbf{F}^* = \begin{array}{c|cccc|c}
& a & b & c & d & \Sigma \\
\hline
a & - & 2 & - & - & 2 \\
b & - & - & 2 & - & 2 \\
c & 1 & - & - & - & 1 \\
d & - & 1 & - & - & 1 \\
\hline
\Sigma & 2 & 2 & 1 & 1 & 6 \\
\end{array}
\]

When we set \(x=2\), \(y=1\), we have the following flow matrix. This is the same as assigning cycle \(\hat{c}_1=abcd\) on top of the premier network:

\[
\mathbf{F} = \begin{array}{c|cccc|c}
& a & b & c & d & \Sigma \\
\hline
a & - & 3 & - & - & 3 \\
b & - & - & 3 & - & 3 \\
c & 1 & - & 2 & - & 3 \\
d & - & 2 & - & - & 2 \\
\hline
\Sigma & 3 & 3 & 2 & 2 & 10 \\
\end{array}
= 2abcd + abc
\]

When we set \(x=3\), \(y=1\), we have the following flow matrix. This is the same as assigning cycle \(\hat{c}_1=abcd\) on top of the premier network twice:

\[
\mathbf{F} = \begin{array}{c|cccc|c}
& a & b & c & d & \Sigma \\
\hline
a & - & 4 & - & - & 4 \\
b & - & - & 4 & - & 4 \\
c & 1 & - & 3 & - & 4 \\
d & - & 3 & - & - & 3 \\
\hline
\Sigma & 4 & 4 & 3 & 3 & 14 \\
\end{array}
= 3abcd + abc
\]

When we set \(x=4\), \(y=1\), we have the following flow matrix. This is the same as assigning cycle \(\hat{c}_1=abcd\) on top of the premier network thrice:

\[
\mathbf{F} = \begin{array}{c|cccc|c}
& a & b & c & d & \Sigma \\
\hline
a & - & 5 & - & - & 5 \\
b & - & - & 5 & - & 5 \\
c & 1 & - & 4 & - & 5 \\
d & - & 4 & - & - & 4 \\
\hline
\Sigma & 5 & 5 & 4 & 4 & 18 \\
\end{array}
= 4abcd + abc
\]

When we set \(x=3\), \(y=2\), we have the following flow matrix. This is the same as assigning cycle \(\hat{c}_1=abcd\) three times and cycle \(\hat{c}_2=abc\) twice:

\[
\mathbf{F} = \begin{array}{c|cccc|c}
& a & b & c & d & \Sigma \\
\hline
a & - & 5 & - & - & 5 \\
b & - & - & 5 & - & 5 \\
c & 2 & - & 3 & - & 5 \\
d & - & 3 & - & - & 3 \\
\hline
\Sigma & 5 & 5 & 3 & 3 & 16 \\
\end{array}
= 3abcd + 2abc
\]

We can see the connection between \(x\) and assigning cycle \(\hat{c}_1=abcd\) on top of the premier network \(x-1\) times. Thus, we can use cycles as variables. Since we have two cycles, we have two variables. We can create a link-cycle matrix where the links are in the rows and the cycles are in the columns and use the linear equation:

\[
\mathbf{H}\mathbf{x} = \mathbf{y}
\]

In our example above:

\[
\mathbf{H} = \begin{array}{c|cc}
& abcd & abc \\
\hline
ab & 1 & 1 \\
bc & 1 & 1 \\
ca & 0 & 1 \\
cd & 1 & 0 \\
da & 1 & 0 \\
\end{array}, \quad \mathbf{x} = \begin{bmatrix}
\alpha_1 \\
\alpha_2 \\
\end{bmatrix},
\]
\[
\mathbf{y} = \begin{bmatrix}

f_{ab} \\
f_{bc} \\
f_{ca} \\
f_{cd} \\
f_{da} \\
\end{bmatrix}
\]

For instance, for Premier Network \[
\mathbf{F}^* = abcd + abc
\] we have \(\mathbf{x} = \begin{bmatrix}
\alpha_1 \\
\alpha_2 \\
\end{bmatrix} = \begin{bmatrix}
1 \\
1 \\
\end{bmatrix}\) produces:

\[
\mathbf{y} = \mathbf{H}\mathbf{x} = \begin{bmatrix}
1 & 1 \\
1 & 1 \\
0 & 1 \\
1 & 0 \\
1 & 0 \\
\end{bmatrix} \cdot \begin{bmatrix}
1 \\
1 \\
\end{bmatrix} = \begin{bmatrix}
2 \\
2 \\
1 \\
1 \\
1 \\
\end{bmatrix}
\]

For for \[
    \mathbf{F} = 4abcd + abc
    \] we can set \(\mathbf{x} = \begin{bmatrix}
\alpha_1 \\
\alpha_2 \\
\end{bmatrix} = \begin{bmatrix}
4 \\
1 \\
\end{bmatrix}\) produces:

\[
\mathbf{y} = \mathbf{H}\mathbf{x} = \begin{bmatrix}
1 & 1 \\
1 & 1 \\
0 & 1 \\
1 & 0 \\
1 & 0 \\
\end{bmatrix} \cdot \begin{bmatrix}
4 \\
1 \\
\end{bmatrix} = \begin{bmatrix}
5 \\
5 \\
1 \\
4 \\
4 \\
\end{bmatrix}
\]

Knowing the premagic flow matrix, we can use the generalized inverse to solve for the number of repetition assignments \(\mathbf{x}\):

\[
\mathbf{x} = \mathbf{H} \backslash \mathbf{y}
\]

For example, for \[
    \mathbf{F} = 3abcd + 2abc
    \] we have

\[
\mathbf{y} = \begin{bmatrix}
5 \\
5 \\
2 \\
3 \\
3 \\
\end{bmatrix}
\]

\[
\mathbf{x} = \mathbf{H} \backslash \mathbf{y} = \begin{bmatrix}
1 & 1 \\
1 & 1 \\
0 & 1 \\
1 & 0 \\
1 & 0 \\
\end{bmatrix} \backslash \begin{bmatrix}
5 \\
5 \\
2 \\
3 \\
3 \\
\end{bmatrix} = \begin{bmatrix}
3 \\
2 \\
\end{bmatrix} = \begin{bmatrix}
\alpha_1 \\
\alpha_2 \\
\end{bmatrix}
\]
\end{exmp}

%% file: StringAnalysis.tex
\section{Signature String Analysis}
In this section, we demonstrate the systematic process of transforming network signatures into practical representations of matrices, vectors, and scalars using only string manipulations.

\begin{defn}
    (Cycle Length): The length of a canonical cycle, denoted \(\left | c_i \right |\), is the number of nodes in the cycle.
\end{defn}

\begin{defn}
    (Kronecker Delta Function):The Kronecker delta function \(\delta\) is defined as:

    \[ \delta_{ij} = \begin{cases}
    1 & \text{if } i = j \\
    0 & \text{if } i \neq j
    \end{cases} \]
\end{defn}

\subsection{Premagic Property}
\begin{lem}\label{lem:premagic from signature}
    (Premagic from Signature): If a matrix \(A\) is derived from a cycle network signature, then \(A\) is a premagic matrix.
\end{lem}
\begin{proof}
Let \(S\) be a cycle network signature consisted of terms \( \alpha_i \hat{c}_i \) where \(\alpha_i\) is the coefficient and \(\hat{c}_i\) is the canonical cycle. Compose the corresponding matrix \(A\) from \(S\), where each entry \(A_{ij}\) represents the flow value between nodes \(i\) and \(j\). For each cycle \(\hat{c}_i\) in \(S\), the flow value \(\alpha_i\) is distributed among the nodes in \(\hat{c}_i\) in such a way that each node receives and sends out an equal amount of flow due to the cyclic nature of \(\hat{c}_i\). For each node \(a\) in \(\hat{c}_i\), the flow contributed to row \(a\) (sum of outgoing flow) is equal to the flow contributed to column \(a\) (sum of incoming flow). Thus, for each node \(a\), the sum of the row entries equals the sum of the column entries, satisfying the condition for a premagic matrix.
\end{proof}

\subsection{Link Flow Value}
\begin{lem}\label{lem:link flow value from signature}
    (Link Flow Value from Signature): The link flow value in the matrix can be obtained from the signature by summing the product of the coefficient with 1 whenever the corresponding link index of the row and the column is in the term.
    \begin{equation}
    f_{pq} = \sum_{i=1}^k \alpha_i \cdot \delta_{pq \in  \hat{c}_i } 
    \end{equation}
\end{lem}
\begin{proof}
    For a given link \(pq\), identify all terms in the signature where the link \(pq\) appears as consecutive nodes. Each occurrence of the link \(pq\) contributes \(\alpha_i\) to the flow value. Use the Kronecker delta function \(\delta_{pq \in \hat{c}_i}\) to represent the presence of \(pq\) in canonical \(\hat{c}_i\). Summing these contributions gives the link flow value for \(pq\). 
\end{proof} 

\begin{corl}
    (Diagonal Entry from Signature): A single letter in the signature corresponds to the entry of a diagonal matrix, meaning:
    \[ A_{ii} = c \quad \text{for some constant } c \]
\end{corl}
\begin{proof}
    Consider a single node \(q\) in the cycle network signature. This corresponds to a diagonal entry in the matrix representation. Let the coefficient of \(q\) be \(c\). Hence, the matrix representation has \(A_{ii} = c\).
\end{proof}

\subsection{Total Flow}
\begin{lem}\label{lem:total flow}
	(Total Flow from Signature): The total flow from a cycle network signature is given by the sum of the product of the coefficient and the cycle length for all terms.
    \begin{equation}\label{eq:total flow}
        \kappa = \sum_{i=1}^k \alpha_i \cdot \left | c_i \right | 
    \end{equation}    
\end{lem}
\begin{proof}
    Let \(\sigma = \sum_{i=1}^k \alpha_i \left | c_i \right |\) where \(\alpha_i\) is the coefficient and \(\left | c_i \right |\) is the cycle length. Each term \(\alpha_i \cdot \left | c_i \right |\) represents the contribution of cycle \(c_i\) to the total flow. Summing these contributions over all terms gives the total flow.
\end{proof}

\subsection{Link Probability}
\begin{lem}\label{lem:Link Probability}
	(Link Probability from Signature): The link probability can be derived by dividing the link flow value by the total flow \(\kappa\).
\end{lem}
\begin{proof}
    From Lemma \ref{lem:link flow value from signature}, the link flow value \(f_{pq}\) can be obtained from the network signature. The total flow \(\kappa\) is given by Lemma \ref{lem:total flow}. The link probability is the ratio of the link flow value to the total flow.
\end{proof}

\subsection{Sum of Rows and Columns}
\begin{lem}\label{lem:Sum of Rows and Columns}
(Sum of Rows/Columns from Signature): The sum of a row or a column in the matrix can be obtained from the signature by summing the product of the coefficient with 1 whenever the corresponding node index of the row or the column is in the term.
\begin{equation}
\sigma^{R}_{q} = \sum_{i=1}^k \alpha_i \cdot \delta_{q \in c_i}
\end{equation}   
\end{lem}
\begin{proof}
For a given node \(q\), identify all terms in the signature where \(q\) appears. Each occurrence of \(q\) contributes \(\alpha_i\) to the sum. Use the Kronecker delta function \(\delta_{q \in c_i}\) to represent the presence of \(q\) in \(c_i\). Summing these contributions gives the row or column sum for node \(q\). Based on Lemma \ref{lem:premagic from signature}, the resulting matrix from network signature is always premagic, the sum of column \(q\) is the same as the sum of row \(q\).
\end{proof}

\subsection{Stochastic Matrices}
\begin{lem}
The row stochastic of outflow probability matrix \(S\) and the column stochastic of inflow probability matrix \(T\) can be derived from network signature.
\begin{equation}    
S_{pq} = \frac{f_{pq}}{\sigma^{R}_{p}} = \frac{\sum_{i=1}^k \alpha_i \cdot \delta_{pq \in c_i}}{\sum_{i=1}^k \alpha_i \cdot \delta_{p \in c_i}} 
\end{equation}
\begin{equation}
T_{pq} = \frac{f_{pq}}{\sigma^{C}_{q}} = \frac{\sum_{i=1}^k \alpha_i \cdot \delta_{pq \in c_i}}{\sum_{i=1}^k \alpha_i \cdot \delta_{q \in c_i}} 
\end{equation}
\end{lem}
\begin{proof}
From the ideal flow matrix, identify the link flow values \(f_{pq}\) using Lemma \ref{lem:link flow value from signature}. Normalize these values by the row sums \(\sigma^{R}_{p}\) for \(S\) and column sums \(\sigma^{C}_{q}\) for \(T\) based on Lemma \ref{lem:Sum of Rows and Columns}.
The stochastic matrix \(S\) represents outflow probabilities, and \(T\) represents inflow probabilities. Thus, the formulations are given by the provided equations.
\end{proof}

\subsection{Pivot}
A pivot is a node that appears in at least two different terms (canonical cycles) in the network signature. Pivot ensures connectivity between the cycles. For two identical cycles, the cycle itself constitutes the pivot. The pivot remains unaffected by the coefficients of the terms.

\begin{defn}
    (Pivot) A \emph{pivot} is defined as the common sequence of nodes between two canonical cycles, which can be a single node, a link (pair of nodes), or a path (sequence of nodes). 
\end{defn}

\begin{lem}\label{lem:Pivot from Signature}
    (Pivot from Signature): Pivot can be obtained from the signature by substring matching between any two cycles. Pivot is a node sequence \( x \) (which can be just one node) found in two canonical cycles.
\begin{equation}
x \in c_i \land x \in c_j \text{ for } i \neq j
\end{equation} 
\end{lem}
\begin{proof}
    Identify common substrings between any two canonical cycles \(c_i\) and \(c_j\). These common substrings represent pivots \(x\). Ensure that \(x\) appears in both cycles \(c_i\) and \(c_j\). Therefore, pivots can be detected through substring matching.
\end{proof} 

%% file: irreducibilityCondition.tex
\subsection{Irreducibility Condition}

A network signature is irreducible if for each term in the signature, there is at least one node (pivot) that overlaps with at least one other term.

\begin{lem}\label{lem-irreducibiliy-signature}
(Irreducibility Condition from Signature) A network signature is designated as \emph{irreducible} if each pair of terms within the signature is connected by at least one pivot. It is not necessary for every pair of terms to have a direct pivot.
\end{lem}

\begin{proof}
Proof by Contradiction: Assume there exists a cycle \(c^*_k\) that does not share any node with any other cycle in the network signature. This would imply the existence of a disconnected subgraph, contradicting the definition of a strongly connected network.
\end{proof}

\begin{exmp}
Network signature \(abcd + cdabe + ef\) is irreducible because:
\begin{itemize}
    \item Between \(abcd\) and \(cdabe\), we find path pivot \(cdab\).
    \item Between \(cdabe\) and \(ef\), we find node pivot \(e\).
    \item There is no need to have a pivot between \(abcd\) and \(ef\).
\end{itemize}
\end{exmp}

\begin{corl}\label{corl-pivot-irreducible}
A matrix is irreducible if there is at least one pivot of any kind between any two cycles in the signature.
\begin{equation}
\forall i \neq j, \exists x \in c_i \land x \in c_j
\end{equation}
\end{corl}
\begin{proof}
Identify pivots \(x\) between cycles \(c_i\) and \(c_j\). Ensure that \(x\) exists in both \(c_i\) and \(c_j\). This guarantees the strong connectivity of the matrix, as every cycle can reach another through pivots. Hence, the matrix is irreducible, as demonstrated by the presence of at least one pivot between any two cycles.		
\end{proof}

The complexity of the Irreducibility Test in Proposition \ref{prop-Matrix-Irreducibility-Test} is \( O(n^3 \log n) \), stemming from matrix multiplication \( O(n^3) \) and \( n-2 \) multiplications. In contrast, the network signature method, with complexity \( O(k^2 n) \), involves \( O(k^2) \) cycle comparisons and \( O(n) \) substring matching. This method can be more efficient, especially when \( k \) (number of cycles) is much smaller than \( n \) (number of nodes), illustrating its advantage in reducing time complexity for testing matrix irreducibility.

\begin{thrm}
An ideal flow matrix can be composed from a network signature if and only if the network signature passes the irreducibility condition.
\end{thrm}

\begin{proof}
    \emph{Sufficient Condition}: Assume that the network signature \(S\) passes the irreducibility condition. By Lemma \ref{lem-irreducibiliy-signature}, a network signature passes the irreducibility condition if every pair of nodes in the corresponding graph is connected by some sequence of edges within \(n-1\) steps. Compose the matrix \(\mathbf{F}\) from the network signature \(S\), where each entry \(F_{ij}\) represents the flow value between nodes \(i\) and \(j\). Since \(S\) passes the irreducibility condition, every node in the matrix \(\mathbf{F}\) is reachable from every other node through a sequence of pivots (common node sequences in the cycles). The presence of these pivots ensures that the matrix \(\mathbf{F}\) is strongly connected by Corollary \ref{corl-pivot-irreducible}, meaning there is a path between any two nodes in the network. Therefore, the matrix \(\mathbf{F}\) is irreducible. From Lemma \ref{lem:premagic from signature}, we know that the matrix \(\mathbf{F}\) is also a premagic matrix, as it is derived from a cycle network signature. By Definition \ref{defn-IFN} an ideal flow matrix is an irreducible premagic matrix. Hence, if the network signature \(S\) passes the irreducibility condition, the matrix \(\mathbf{F}\) derived from \(S\) is an ideal flow matrix.
    \emph{Necessary Condition}: Assume that \(\mathbf{F}\) is an ideal flow matrix derived from a network signature \(S\). By Definition \ref{defn-IFN}, an ideal flow matrix is an irreducible premagic matrix. From the definition of irreducibility, for \(\mathbf{F}\) to be irreducible, there must be a path between any two nodes within \(n-1\) steps. This implies that the network signature \(S\) must have pivots (common node sequences in the cycles) ensuring strong connectivity between all nodes. Decompose the network signature \(S\) such that it represents the cycle structure and flow values of the matrix \(\mathbf{F}\). Since \(\mathbf{F}\) is irreducible, the network signature \(S\) must pass the irreducibility condition, meaning that every pair of nodes in the corresponding graph is connected by some sequence of edges within \(n-1\) steps. Therefore, if \(\mathbf{F}\) is an ideal flow matrix, the network signature \(S\) from which it is derived must pass the irreducibility condition. Hence, an ideal flow matrix can be composed from a network signature if and only if the network signature passes the irreducibility condition.
\end{proof}

\subsection{Random IFN}
\begin{lem}
    (Random IFN from Signature): A random integer ideal flow matrix of a certain node \( N \) and a certain total flow \(\kappa\) can be obtained by creating random canonical cycles with a number of nodes less than or equal to \( N \) and ensuring the existence of pivots between consecutive terms.
\end{lem}
\begin{proof}
Start with an initial cycle of length \(N\). Assign a coefficient such that the total flow \(\kappa\) is evenly distributed. Create subsequent cycles by introducing random pivots and ensuring connectivity. Adjust coefficients to maintain the total flow \(\kappa\), solving the linear Diophantine equation (\ref{eq:total flow}) on total flow from signature.
\end{proof}

%% file: conclusion.tex
\section{Conclusion}

The introduction of Ideal Flow Networks (IFNs) and their associated signatures offers a novel approach to network theory, with significant implications for both theoretical exploration and practical application in network analysis. This paper establishes a robust framework for the composition and decomposition of IFNs, leveraging canonical cycles and network signatures. We elucidate their properties and their intrinsic relationships with matrices and network counterparts. The theory underscores the utility of linear algebra in managing network flows, demonstrating the efficiency and minimality of network signatures for algorithmic implementations. This framework not only validates the theoretical constructs but also opens new pathways for research and application across various disciplines.

%% file: supplementary.tex
\section*{Supplementary Materials}
An interactive online program has been developed to aid in the exploration of the framework described in this paper. Readers can access this tool at \href{https://people.revoledu.com/kardi/tutorial/IFN/CompositionDecomposition.html}{IFN Playground} to engage more deeply with the presented concepts and theorems.